\documentclass[%
 reprint,
superscriptaddress,
nofootinbib,
 amsmath,amssymb,
 aps,
prb,
]{revtex4-1}
\usepackage[section]{placeins}
\usepackage{graphicx}% Include figure files
\usepackage{dcolumn}% Align table columns on decimal point
\usepackage{bm}% bold math
\usepackage[version=3]{mhchem}
\usepackage{color}
\usepackage{float}

\begin{document}
\sloppy
\preprint{APS/123-QED}

\title{Conventional superconductivity and charge-density-wave ordering in \ce{Ba_{1-\textit{x}}Na_{\textit{x}}Ti2Sb2O}}

\author{Fabian von Rohr}
\email{vonrohr@physik.uzh.ch}
\affiliation{Physik-Institut der Universit\"at Z\"urich, Winterthurerstrasse 190, CH-8057 Z\"urich, Switzerland}
\affiliation{Lab. of Inorg. Chemistry, ETH Z\"urich, Wolfgang-Pauli-Str. 10, CH-8093 Z\"urich, Switzerland}

\author{Andreas Schilling}
\affiliation{Physik-Institut der Universit\"at of Z\"urich, Winterthurerstrasse 190, CH-8057 Z\"urich, Switzerland}

\author{Reinhard Nesper}
\affiliation{Lab. of Inorg. Chemistry, ETH Z\"urich, Wolfgang-Pauli-Str. 10, CH-8093 Z\"urich, Switzerland}

\author{Chris Baines}
\affiliation{Laboratory for Muon Spin Spectroscopy, Paul Scherrer Institut, CH-5232 Villigen PSI, Switzerland}

\author{Markus Bendele}
\affiliation{Dipartimento di Fisica, Università di Roma "La Sapienza", P.le Aldo Moro 2, 00185 Roma, Italy}

\date{\today}% It is always \today, today,

\begin{abstract} We have investigated the low temperature physical properties of \ce{BaTi2Sb2O} and \ce{Ba_{1-\textit{x}}Na_{\textit{x}}Ti2Sb2O} ($x$ = 0.05, 0.1, 0.15, 0.2, 0.25, 0.3) by means of muon spin rotation ($\mu$SR) and SQUID magnetometry. Our measurements reveal the absence of magnetic ordering below $T_{DW}$ = 58 K in the parent compound. Therefore the phase transition at this temperature observed by magnetometry is most likely due to the formation of a charge density wave (CDW). Upon substitution of barium by sodium in \ce{Ba_{1-\textit{x}}Na_{\textit{x}}Ti2Sb2O} we find for $x = 0.25$ superconductivity with a maximum $T_c$ = \nolinebreak[4]  5.1 K in the magnetization and a bulk $T_{c,bulk}$ = 4.5 K in the $\mu$SR measurements. The temperature dependency of the London penetration depth $\lambda^{-2}(T)$ of the optimally doped compound can be well explained within a conventional weak-coupling scenario in the clean limit.
\end{abstract} 

\maketitle

%\tableofcontents

\section{Introduction}
Nesting at the Fermi-surface is known to be a key feature for the occurrence of either charge (CDW) or spin density wave (SDW) ordering, and it is considered to be of importance for the emergence of superconductivity in some materials (e.g. \ce{BaFe_{2-x}Co_xAs_2} \cite{iron_nature}). The competition or coexistence of superconductivity and spin density wave (SDW) ordering is one of the most extensively discussed topics for iron-based superconductors and for the stripe phases of cuprates.\cite{Luetkens, stripe_phase} The competition or coexistence between superconductivity and CDW ordering at low temperatures is less often encountered (see e.g. in \ce{Cu_xTiSe2} \cite{CuxTiSe2}, in 2H-NbSe$_2$ \cite{NbSe2}, and in \ce{Ba_{1-\textit{x}}K_{\textit{x}}BiO3} \cite{Cox,Hamann}), though the development of CDW order at zero field in the normal state of superconducting \ce{YBa2Cu3O_{6.67}} has been prominently discussed.\cite{YBCO_DW} \\

The large family of stacked, layered titanium oxide pnictide compounds were long considered as potential host structures for superconductivity. \cite{Johrendt} Most of these materials were identified to undergo magnetic or density wave (DW) ordering transitions at low temperatures, in the absence of superconductivity. For example, $\mathrm{Na_2Ti_2As_2O}$ and $\mathrm{Na_2Ti_2Sb_2O}$, which crystallize in a modified anti-\ce{K2NiF4} type structure, were found to undergo transitions to SDW ordered states at $T_{SDW}$ of 320 K and 115 K, respectively. \cite{Na2Ti2Sb2O} \\

\ce{BaTi2Sb2O} belongs to this family of compounds. Its structure consists of titanium, octahedrally surrounded by oxygen, leading to square planar \ce{Ti2O} sheets. \cite{ZAAC} This compound was found to undergo a phase transition to a spin or charge density wave (SDW or CDW) around $T_{DW}$ = 55 K and it was proposed that below $T_c$ = 1 K it is a superconductor. \cite{BaTi2Sb2O} Recently, it was shown that upon substitution of barium by sodium in \ce{Ba_{1-\textit{x}}Na_{\textit{x}}Ti2Sb2O}, $T_{DW}$ is lowered and eventually suppressed, while superconductivity reaches a maximum $T_c$ of approximately 5 K. \cite{Doan_JACS} \\

In this article we will show by a series of SQUID magnetometry and muon-spin rotation ($\mu$SR) experiments that \ce{Ba_{1-\textit{x}}Na_{\textit{x}}Ti2Sb2O} is another example for the coexistence and competition of the periodic modulations of CDW and superconductivity. Our results suggest that the CDW ordering competes with a conventional superconducting state in these materials.

\section{Experimental}
Standard solid-state reactions were employed to synthesize polycrystalline samples of \ce{Ba_{1-\textit{x}}Na_{\textit{x}}Ti2Sb2O} with $x$ = 0, 0.05, 0.10, 0.15, 0.20, 0.25 and 0.30. BaO (99.99\%), \ce{BaO2} (95\%), \ce{Na2O2} (95\%), Ti (99.99\%) and Sb (99.999\%) were mixed and pressed into pellets in an argon filled glove box. The pellets were sealed in argon filled niobium ampules and then sintered at 1000 \nolinebreak[4] $^{\circ}$C for 24 h. Then the samples were reground under inert atmosphere, repelletized and sintered again for 36 h at 1000 \nolinebreak[4] $^{\circ}$C. The purity, symmetry, and cell parameters were checked by x-ray powder diffraction using a Stoe STADIP diffractometer (Cu-$\mathrm{K_{\alpha 1}}$ radiation, $\lambda$ = 1.54051 \nolinebreak[4] $\mathrm{\AA}$, Ge-monochromator).\\ 

\begin{figure} [t!]
\centering
%{\includegraphics[width=\textwidth]{images/fig1}}
\includegraphics[width=0.85\linewidth]{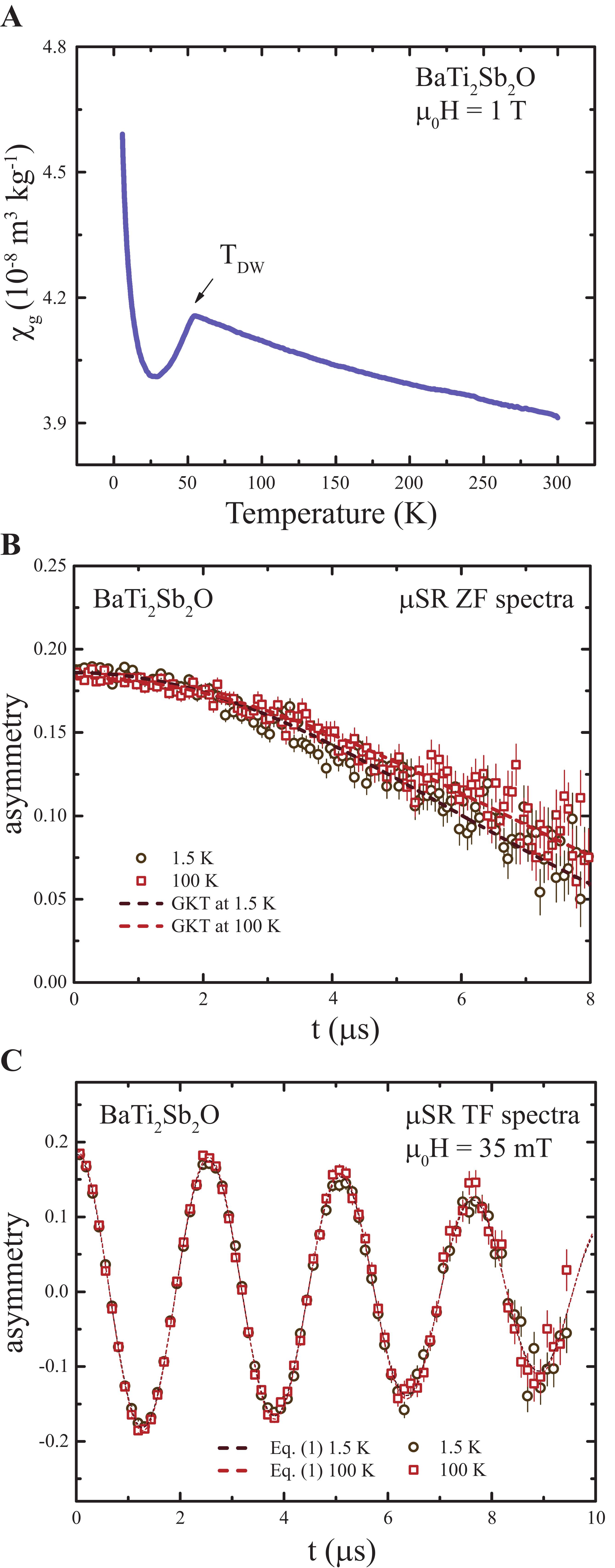}
\caption{Parent compound, \ce{BaTi2Sb2O}, characterized by (A) the temperature-dependent susceptibility in a field of $\mu_0H$ = 1 T, %(B) the normalized field-dependent magnetization in fields of $\mu_0H$ = 0 T to 7 T at temperatures T = 30 K and 100 K, 
(B) the ZF $\mu$SR spectra at 1.5 K and 100 K (the dashed lines are fits to the Gaussian Kubo-Toyabe function) and (C) TF $\mu$SR spectra (the dashed lines are fits to equation (\ref{Eq_raw}) at temperatures T = 1.5 K and 100 K).}
\label{fig:parent_compound}
\end{figure}

The magnetic properties were studied using a Quantum Design Magnetic Properties Measurements System (MPMS XL) equipped with a reciprocating sample option (RSO). Transverse field (TF) and zero field (ZF) $\mu$SR experiments were carried out at the LTF instrument at the $\pi$M3 beamline, and at the Dolly instrument at the $\pi$E1 beamline at the Paul Scherrer Institute (PSI), Switzerland. The superconducting pellets were cooled from above $T_c$ in a field of $\mu_0H$ = 35 mT for the TF experiments. The occurrence of magnetism was investigated in these samples with the ZF experiments. The $\mu$SR time spectra have been analyzed using the free software package MUSRFIT. \cite{MUSRFIT}

\section{Results and Discussion}
In figure \ref{fig:parent_compound}a we show magnetization $M(T)$ data of the parent compound \ce{BaTi2Sb2O}, in a field of $\mu_0 H$ = 1.0 T, showing a distinct kink at $T_{DW}$ = 58 K. This discontinuity was earlier attributed to either a SDW or a CDW ordering transition. \cite{BaTi2Sb2O} \\ 

The ZF and weak TF muon time signals for \ce{BaTi2Sb2O} were measured above and below $T_{DW}$ at $T$ = 1.5 K and $T$ = 100 K, as shown in figure \ref{fig:parent_compound}b and \ref{fig:parent_compound}c. The measurements show neither indications of static nor fluctuating magnetism down to $T=1.5$ K. Moreover, the relaxation rates are small and show only little differences between the measurements at high temperatures and at 1.5 K. The ZF spectra are well described by a standard Gaussian Kubo-Toyabe (GKT) function \cite{GKT}, which is typical for nuclear moments. The zero field $\mu$SR spectra above and below $T_{DW}$ do not exhibit any noticeable change in the relaxation rate, indicating the absence of a spontaneous internal field at the muon stopping site (within the sensitivity of $\mu$SR). This is further supported by the weak TF measurements (Fig. \ref{fig:parent_compound}c), where no reduction of the asymmetry is observed, as it would be expected in case of magnetic ordering. Therefore, we can exclude that the observed transition at $T_{DW}$ is caused by SDW ordering in the parent compound BaTi$_2$Sb$_2$O. The observed transition is therefore most likely caused by CDW ordering. These findings are in agreement with recent NMR measurements. \cite{NMR}\\

\begin{figure}
\centering
\includegraphics[width=0.85\linewidth]{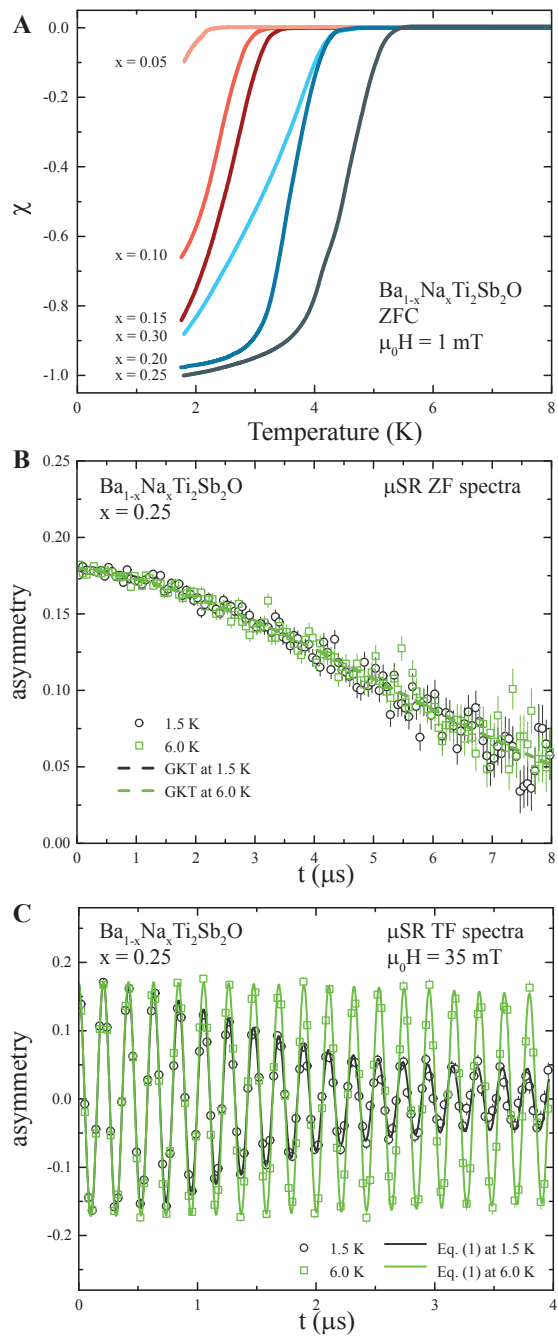}
\caption{(A) The magnetic susceptibility of \ce{Ba_{1-\textit{x}}Na_{\textit{x}}Ti2Sb2O} for $x$ = 0.05, 0.1, 0.15, 0.2, 0.25, 0.3 measured in a field of $\mu_0$H = 1 mT. (B) The ZF $\mu$SR spectra for $x$ = 0.25 above (6 K) and below (1.5 K) $T_c$ (the dashed lines are the fits to the Gaussian Kubo-Toyabe function). (C) The TF $\mu$SR spectra for $x$ = 0.25 above (6 K) and below (1.5 K) $T_c$ (the solid lines are fits to equation (\ref{Eq_raw})). The strong relaxation of the signal at 1.5 K can be ascribed to the presence of the flux-line lattice.}
\label{fig:supercond}
\end{figure}

Upon substitution of barium by sodium in \ce{Ba_{1-\textit{x}}Na_{\textit{x}}Ti2Sb2O}, we find superconductivity with a maximum $T_c$ = 5.1 K in the magnetization, and a bulk $T_{c,bulk}$ = 4.5 K in the $\mu$SR measurements, for $x = 0.25$. The temperature dependent measurements of the DC magnetic susceptibility in the vicinity to superconductivity (1.8 K to 10 K), measured in zero-field cooled (ZFC) mode in an external field of $\mu_0 H$ = 1 mT, are shown in figure \ref{fig:supercond}a. The transitions to the superconducting state are depicted for six representative members of the series, $x$ = 0.05, 0.1, 0.15, 0.2, 0.25 and 0.3. In figure \ref{fig:supercond}b we show the ZF muon time signals for the optimally doped sample $x$ = 0.25 at $T=1.5$ K and above $T_c$ (6 K). The ZF spectra are well described by a GKT function and are overlapping for both measurements, revealing no magnetic ordering down to 1.5 K. The relaxation above $T_c$ in the TF measurements in a field of $\mu_0H$ = 35 mT is shown in figure \ref{fig:supercond}c. Similar to the ZF measurements (figure \ref{fig:supercond}b), only a small relaxation arising from the randomly aligned nuclear magnetic moments is observed. The strong additional relaxation in the TF measurements below $T_c$, however, is solely due to the formation of the flux-line lattice (FLL) in the Shubnikov phase. As shown by Brandt, the second moment of the resulting inhomogeneous field distribution is related to the magnetic penetration depth $\lambda$ as $\left<\Delta B^2\right>\propto  \sigma^2_{sc} \propto \lambda^{-4}$, whereas $\sigma_{sc}$ is the Gaussian relaxation rate due to the formation of the FLL. \cite{Brandt88,Brandt03} The TF $\mu$SR time evolutions were analyzed using the following functional form for the polarization:
\begin{equation}
 A(t)=A(0)\exp \left[-\frac{\sigma_{sc}^2+\sigma_{nm}^2}{2}t^2\right]\cos (\gamma_\mu B_{int}t+\varphi)+A_{BG}.
\label{Eq_raw}
\end{equation}
Here, $A(0)$ and $\varphi$ are the initial asymmetry and the phase of the muon ensemble, respectively, $\sigma_{nm}$ is the damping arising from the nuclear magnetic dipole moments, which we assumed to be temperature independent and fixed to the value obtained above $T_c$, $\gamma_\mu/(2\pi)=135.5$ MHz/T is the muon gyromagnetic ratio, and $B_{int}$ represents the internal magnetic field at the muon stopping site. For the low temperature measurements in the LTF instrument, part of the muon beam is stopped in the silver sample holder, resulting in a background denoted as $A_{BG}$.\\

\begin{figure}
\centering
%\vspace{-0cm}
\includegraphics[width=1\linewidth]{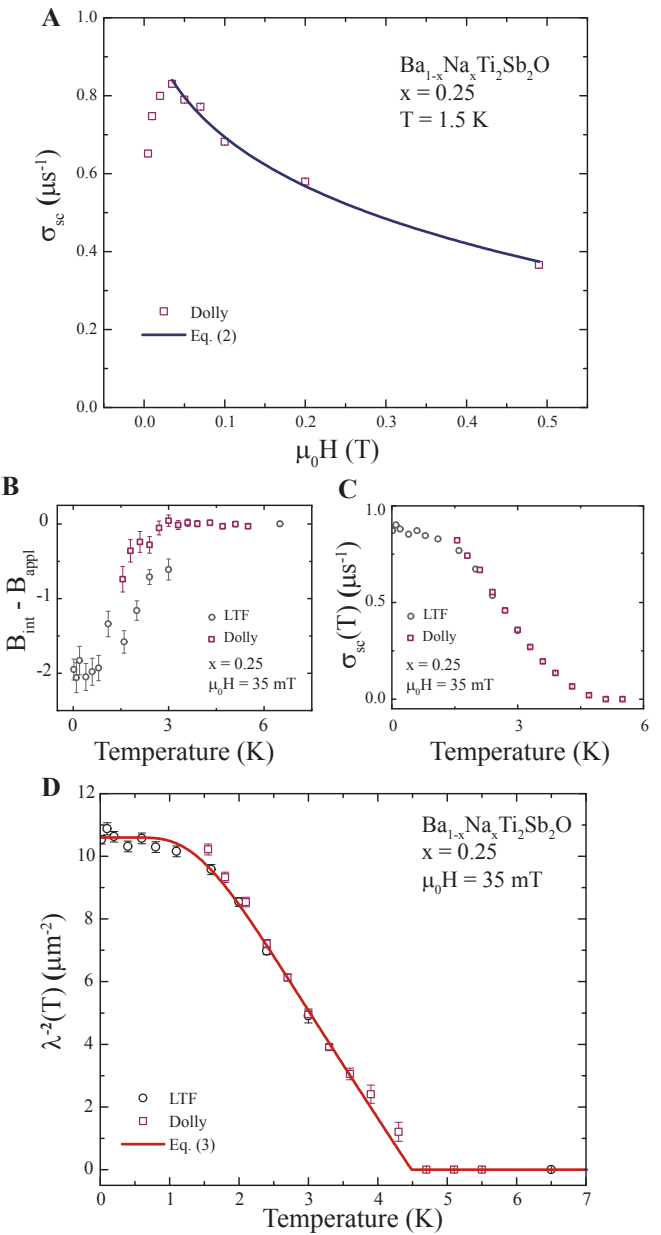}
\caption{(A) Field dependence of the muon depolarization rate $\sigma_{sc}$ at $T\simeq1.5$ K for the optimal doping $x=0.25$. The solid line corresponds to a fit of the experimental data to equation (\ref{Eq_field}). The insets show the corresponding field dependence of $\lambda^{-2}$ (in the Shubnikov phase). (B) The diamagnetic field shift in the superconducting state with respect to above $T_c$ ($B_{int}-B_{appl}$) for $x=0.25$.  (C) The temperature dependence of the muon polarization rate $\sigma_{sc}(T)$ measured in $\mu_0H=35$ mT. (D) The temperature dependence of $\lambda^{-2}$ for $x = 0.2$ as reconstructed from $\sigma_{sc}(T)$ (shown in B), measured in $\mu_0H=35$ mT. The solid line corresponds to a fit to equation (\ref{Eq_swave}) with $2 \Delta/(k_B T_c)$ = 2.9. Squares: Dolly instrument, circles: LTF instrument.}
\label{fig:muSR}
\end{figure}

For a weak coupling BCS superconductor and $B_{ext} \ll B_{c2}$, $\lambda$ does not depend on external magnetic fields, whereas e.g. in a multiple gap or nodal superconductor $\lambda$ can be significantly field dependent. \cite{Bendele10,Khasanov07,Sonier00,Serventi04} In case of an ideal vortex lattice of an isotropic $s$-wave superconductor within the Ginzburg-Landau theory, the relaxation rate $\sigma^2$ in the superconducting state should follow the expression \cite{Brandt03}
\begin{equation}
 \sigma_{sc}=a\cdot\left(1-\frac{B}{B_{c2}}\right)\left[1+1.21\left(1-\sqrt{\frac{B}{B_{c2}}}\right)^3\right]\lambda^{-2}.
\label{Eq_field}
\end{equation}
Here, $a$ is a coefficient given by the symmetry of the vortex lattice (with $a = 4.83 \cdot10^4 \ nm^2/\mu sec$ for triangular and $a = 5.07 \cdot10^4 \ \mathrm{nm^2/\mu sec}$ for a rectangular vortex lattice geometry \cite{numericalVL,Brandt03}), $B$ is the magnetic induction, for which we may assume $B\simeq B_{ext}$ in the region $\mu_0 H_{c1}\ll B_{ext}\ll \mu_0 H_{c2}$ (with $H_{c1}$ the lower and $H_{c2}$ the upper critical field, respectively). Equation (\ref{Eq_field}) in general accounts for the reduction of $\sigma_{sc}$ due to the stronger overlap of the vortices with increasing field. A fit of the measured $\sigma_{sc}$ according to equation (\ref{Eq_field}) describes the data reasonably well and yields $B_{c2} = 1.5(1)$ T and $\lambda^{-2}=10^{-5}(0.01)$ nm$^{-2}$ at $T=1.5$ K for $x=0.25$ (see figure \ref{fig:muSR}a). The value of $B_{c2}$, for the optimal doping $x=0.25$, is in excellent agreement with previous measurements.\cite{Gooch13} The parameter $a$ was fitted to $4.87(5)\cdot10^4 \ \mathrm{nm^2/\mu sec}$ indicating that the vortex lattice in Ba$_{1-x}$Na$_x$Ti$_2$Sb$_2$O has triangular shape. To obtain maximum field contrast, we chose the magnetic field $\mu_0H=$ 35 mT to study the temperature dependence of $\lambda(T)$ for the optimally doped sample $x$ = 0.25. Measurements down to $T=0.02$ K and $T=1.5$ K were performed in the LTF and Dolly instruments, respectively. A diamagnetic shift of the internal magnetic field $B_{int}$ is observed below $T_c$ (figure \ref{fig:muSR}b). The resulting temperature dependence of $\sigma_{sc}$ is shown in figure \ref{fig:muSR}c. In figure \ref{fig:muSR}d we show the temperature dependence of $\lambda^{-2}(T)$ as reconstructed from $\sigma_{sc}(T)$, using equation (\ref{Eq_field}). The temperature dependence of $B_{c2}(T)$, used in the corresponding calculation according to equation 2, was assumed to follow the theoretical Werthamer-Helfand-Hohenberg relation. \cite{Gooch13,WHH} \\

These measurements suggest that $\lambda^{-2}$ is virtually temperature independent below $T\simeq 1$ K for the optimally doped sample. The obtained experimental temperature dependence of $\lambda^{-2}(T)$ was tentatively analyzed within the clean limit approach for a London superconductor with an $s$-wave gap\cite{Tinkham}
\begin{equation}
\frac{\lambda^{-2}(T)}{\lambda^{-2}(0)}=1+2\int_{\Delta(T)}^{\infty}\left( \frac{\partial f}{\partial E} \right)\frac{E}{\sqrt{E^2-\Delta^2(T)}}dE.
\label{Eq_swave}
\end{equation}
Here $\lambda(0)$ is the zero temperature value of the magnetic penetration depth, $f=[1+\exp(E/k_BT)]^{-1}$ is the Fermi function (with $k_B$ the Boltzmann constant), and $\Delta(T)=\Delta(0)\tilde{\Delta}(T/T_c)$ represents the temperature dependence of the energy gap, which can be approximated to sufficient precision as $\tilde{\Delta}(T/T_c)=\tanh(1.82[1.018(T_c/T-1)^{0.51}])$.\cite{Carrington03} The results of this fit are $T_c=4.49(6)$ K and $\Delta(0)=0.56(1)$ meV with a zero temperature magnetic penetration depth $\lambda=307(10)$ nm. This corresponds to a ratio $2\Delta/(k_BT_c)=2.9$, which is quite close to the value of a weak coupling BCS superconductor $2 \Delta/(k_B T_c)$ = 3.5. There are no signs of multi-gap superconductivity in these data (compare Ref. \onlinecite{Bendele10,Khasanov07,Sonier00,Serventi04}), and the presented low temperature $\lambda^{-2}(T)$ data taken in a low magnetic field seem to be incompatible with a possible $d$-wave scenario \cite{Amin00}.

\section{Conclusion}
We have presented magnetization and $\mu$SR results on the density wave (DW) ordering transition in \ce{BaTi2Sb2O}, and on the transition to superconductivity in \ce{Ba_{1-\textit{x}}Na_{\textit{x}}Ti2Sb2O}. The observed absence of a magnetic contribution to the $\mu$SR data related to the phase transition at $T_{DW}$ = 58 K of the parent compound \ce{BaTi2Sb2O} is strong evidence against the in Ref. \onlinecite{Singh} theoretically proposed SDW ordering transition. Therefore the observed phase transition is most likely due to CDW ordering that competes with a superconducting state. 
Upon substitution of barium by sodium in \ce{Ba_{1-x}Na_xTi2Sb2O} we find superconductivity with a maximum $T_c$ = 5.1 K in the magnetization, and a bulk $T_{c,bulk}$ = 4.5 K in the $\mu$SR measurements, for $x = 0.25$. In the TF $\mu$SR spectra for $x$ = 0.25 below $T_c$ a strong relaxation of the signal is observed, which is due to the formation of the flux-line lattice. This is strong evidence for the bulk nature of the superconductivity in this material. The obtained experimental temperature dependence of $\lambda^{-2}(T)$ can be reasonably well explained within the clean limit approach for a conventional London superconductor, which is consistent with recently published NMR and specific heat results, as well as theoretical calculations. \cite{NMR,Gooch13,Subedi}
\section*{Acknowledgments}
FvR acknowledges a scholarship from Forschungskredit UZH, grant no. 57161402. MB acknowledges the financial support of the Swiss National Science Foundation (grant number PBZHP2143495).

%\bibliography{BaTi2Sb2O}

%\bibliographystyle{plain}
 %\bibpunct{}{}{;}{n}{,}{,}
% \newcounter{ser}

%\bibliographystyle{unsrt}

 \end{document}